\documentclass[11pt,preprint]{aastex}
\usepackage{color}
\usepackage{lscape}

\begin{document}

\shorttitle{}
\shortauthors{Ferraro  et al.}

\title{The age of the young bulge-like population in the stellar system
  Terzan5: linking the Galactic bulge to the high-z
  Universe\footnotemark[1]}

\footnotetext[1]{Based on data obtained with (1) the ESA/NASA HST, under programs 
GO-14061, GO-12933, GO-10845, (2) the Very Large Telescope of 
the European Southern Observatory during the Science Verification of the camera MAD; 
(3) the W.M. Keck Observatory, which is operated as a scientific 
partnership among the California Institute of Technology, 
the University of California and NASA.}

\author{
F. R. Ferraro\altaffilmark{2},
D. Massari\altaffilmark{3,4},
E. Dalessandro\altaffilmark{2,3},
B. Lanzoni\altaffilmark{2},
L. Origlia\altaffilmark{3},
R. M. Rich\altaffilmark{5},
A. Mucciarelli\altaffilmark{2}
}
\affil{\altaffilmark{2} Dipartimento di Fisica e Astronomia, Universit\`a degli
  Studi di Bologna, Viale Berti Pichat 6/2, I--40127 Bologna, Italy}
\affil{\altaffilmark{3} INAF- Osservatorio Astronomico di Bologna, Via
  Ranzani, 1, 40127 Bologna, Italy}  
\affil{\altaffilmark{4} Kapteyn Astronomical Institute, University of Gr$\ddot{\rm o}$ningen,  
 Kapteyn Astron Institute, NL-9747 AD Gr$\ddot{\rm o}$ningen, Netherlands}
\affil{\altaffilmark{5} Department of Physics and Astronomy, 430 Portola Plaza, 
Box 951547, Los Angeles, CA 90095-1547, USA}
  
\date{5 Sept 2016}

\begin{abstract}
The Galactic bulge is dominated by an old, metal rich stellar
population.  The possible presence and the amount of a young (a few
Gyr old) minor component is one of the major issues debated in the
literature. Recently, the bulge stellar system Terzan 5 was found to
harbor three sub-populations with iron content varying by more than
one order of magnitude (from 0.2 up to 2 times the solar value), with
chemical abundance patterns strikingly similar to those observed in
bulge field stars. Here we report on the detection of two distinct
main sequence turn-off points in Terzan 5, providing the age of the two
main stellar populations: 12 Gyr for the (dominant) sub-solar
component and 4.5 Gyr for the component at super-solar
metallicity.  This discovery classifies Terzan 5 as a site in the
Galactic bulge where multiple bursts of star formation occurred, thus
suggesting a quite massive progenitor possibly resembling the giant
clumps observed in star forming galaxies at high redshifts. This
connection opens a new route of investigation into the formation
process and evolution of spheroids and their stellar content.
\end{abstract}


\section{Introduction}
\label{intro}
The picture of galaxy bulges formation is still highly debated in the
literature \citep[for a review of the Milky Way bulge, see,
  e.g.,][]{rich13,origlia14}.  Among the proposed scenarios, three main channels
can be grossly distinguished: dissipative collapse
\citep[e.g.,][]{ballero07, mcwilliam08}, with possibly an additional
component formed with a time delay of a few Gyr \citep[e.g.][]{tsu12,
  grieco12}, dynamical secular evolution of a massive disk that
buckles into a bar \citep[e.g.,][]{combes81, raha91, saha13}, and
merging of substructures either of primordial galaxies embedded in a
dark matter halo \citep[e.g.,][]{scannapieco03, hopkins10}, or massive
clumps generated by early disk fragmentation \citep[e.g.][]{immeli04,
  carollo07, elmegreen08}.  In the merging scenarios most
of the early fragments rapidly dissolved/merged together to form the
bulge. However, a few of them could have survived the total disruption
\citep[e.g.][]{immeli04, carollo07, elmegreen08} and it is very
possible that such fossil relics are still observable somewhere in the
host galaxy, grossly appearing as normal globular clusters
(GCs). Because of their original large mass, these clumps should have
been able to retain the iron-enriched ejecta and the stellar remnants
of the supernova (SN) explosions, and they likely experienced more
than one burst of star formation. As a consequence, we expect these
fossil relics to host multi-\emph{iron} sub-populations and, possibly,
also a large number of neutron stars.  Clearly, finding stellar
systems with these properties would provide crucial observational
support to the 'merging" scenario for bulge formation.

Until recently, no empirical probes of such fossil clumps in galaxy
bulges were available. The situation changed in 2009, when
\citet{ferraro09} discovered two stellar components with very
different \emph{iron} abundances in Terzan 5, a stellar system in the
Milky Way bulge previously catalogued as a GC, with the only
peculiarity of hosting the largest population of millisecond pulsars
in the Galaxy \citep{ransom05}.  The two components of Terzan 5 appear
well distinct at the level of the red clump (RC) and red giant branch
(RGB) in the combined NIR-optical color-magnitude diagram (CMD;
\citealp{ferraro09, massari12}).  Moreover, they display significantly
different iron and $\alpha$-element abundances, the metal-poor
population having [Fe/H]$=-0.25$ dex and [$\alpha/$Fe]$=+0.34$, the
metal-rich one showing [Fe/H]$=+0.27$ dex and [$\alpha/$Fe]$=+0.03$
(\citealp{origlia11}; a minor component at [Fe/H]$=-0.8$ has been also
detected, extending the internal metallicity range of Terzan 5 over
more than 1 dex; see \citealp{origlia13, massari14}).  Given these
abundance patterns, the observed RC split could be explained either in
terms of an age difference of a few Gyr, or in terms of a different
helium content in two nearly coeval sub-populations \citep{dantona10,
  lee15}.
 
The chemical abundance patterns measured in Terzan 5 are strikingly
similar to those observed  toward the Galactic bulge, while no
other stellar system within the Milky Way  outer disk and halo or
in the Local Group show analogous properties \citep{chiappini99,
  matteucci05, lemasle12}.  This opens the intriguing possibility that
Terzan 5 is a fossil relic of one of the structures that contributed
to generate the Galactic bulge.  The bulge is known to be dominated by
an old ($>10$ Gyr) stellar population with solar-like metallicity
\citep{clarkson08,valenti13,rich13}. However, growing indirect
evidence \citep{bensby13, dekany15, nataf15} supports the existence of
a minor, significantly younger component, with age not precisely
determined yet. Thus, measuring the absolute ages of the two major
stellar components in Terzan 5 is of paramount importance also in the
context of the bulge formation history.

Here we present the accurate determination of the main sequence
turn-off (MS-TO) region in Terzan 5, from which we determined the age
of its two main stellar populations.  The paper is organized as
follows. The data-sets used and the data reduction procedures are
presented and discussed in Section 2.  The analysis of the CMDs, and
the measure of the ages and radial distributions of the two
populations are presented in Section 3. Section 4 is devoted to
discussion and conclusions.


\section{Observations and Data Analysis}
\label{obs}
For this study we used a wide photometric database collected with
different instruments and telescopes: (i) ultra-deep images in the
F606W and F814W filters acquired with the Advanced Camera for Surveys
(ACS) on board the Hubble Space Telescope (HST), in three different
epochs spanning more than 11 years (GO: 9799, PI: Rich; GO: 12933, PI:
Ferraro and GO: 14061, PI: Ferraro); (ii) a set of images in the F110W
and F160W filters secured with the IR channel of the Wide Field Camera
3 (WFC3) on board the HST (GO: 12933, PI: Ferraro); (iii) a set of
high-resolution images in the $J$ and $K$ bands acquired with the
Multi-conjugate Adaptive Optic Demonstrator (MAD) at the ESO Very
Large Telescope (Science Demonstration Proposal, PI: Ferraro); (iv) a
set of $K$-band images obtained at the Keck Telescope by using the
camera NIRC2 assisted with laser adaptive optics (U156N2L, PI: Rich).
Most of the dataset used in this work was already presented and
discussed in previous papers
\citep[see][]{ferraro09,lanzoni10,massari14,massari15,ferraro15}. Hence
detailed information on the observations, data quality and data
analysis can be found in those papers, while here we report only a
schematic summary of the datasets and the reduction processes.
 
The optical HST dataset acquired with the Wide Field Channel (WFC) of
the ACS camera is described in detail in \citet{massari15} and 
\citet{ferraro15}. Here we used exclusively the \_FLC images, already flat
fielded and corrected for charge transfer efficiency (CTE) losses with
the pixel-based correction in the pipeline \citep{jaybedin10,
  ubedajay}. The data reduction was performed using the software
DAOPHOT-II (\citealt{stetson87, stetson94}), following the standard
procedure described e.g. in \citet{dalessandro15}. Briefly, for each
exposure we determined the best PSF model using a few hundreds of
bright, isolated stars. The model was then applied to all the sources
detected (using ALLSTAR) above a 3$\sigma$ threshold from the
background.  Since the images secured with the F814W filter (being
less affected by the strong extinction present in the cluster
direction) turned out to be significantly deeper than those acquired
in the filter F606W, we created a master list of stars composed by
sources detected in at least three frames.  This master list was used
as input to force the PSF fitting with DAOPHOTII/ALLFRAME. The star
magnitudes obtained in each frame secured in the same filter were then
homogenized, and their weighted mean and standard deviation were
finally adopted as magnitude and photometric error of each source. The
stellar instrumental magnitudes and positions were finally calibrated
onto the Johnson-Cousins system and reported onto the 2MASS equatorial
reference frame, respectively, as described in \cite{lanzoni10}.

The WFC3 dataset is made of 32 images, each one acquired with an
exposure time of 300 s in both the F110W and the F160W filters. We
worked on the images already pre-reduced by the STScI pipeline,
following the same reduction procedure described in the previous
paragraph. Also in this case, the F814W ACS catalog was used as input
for the ALLFRAME analysis of all the exposures. The final catalog was
built with the same prescriptions used for the ACS one. The
instrumental magnitude were calibrated using the aperture correction
and the zero-points listed at
\texttt{http://www.stsci.edu/hst/wfc3/phot$\_$zp$\_$lbn}.  Stellar
positions were brought to the 2MASS astrometric reference using the
CataPack suite of software\footnote{{\tt
    www.bo.astro.it/$\sim$paolo/Main/CataPack.html}}

The MAD dataset is already described in \cite{ferraro09}. Since the
observations span a temporal interval of several hours, with seeing
conditions varying quite significantly from one exposure to another,
we limited our analysis only to those with a stable Full Width at Half
Maximum (FWHM) of $\sim 0.1\arcsec$ across the entire MAD field of
view. FWHM were measured on bright, non saturated and isolated stars
using the standard IRAF tools. Fifteen exposures survived such a
selection, and we reduced them as described for the previous data
sets. In this case, however, the PSF model was computed by allowing
the look-up table of the star profile fitting residuals to vary
across the field of view as a third order polynomial. This is because,
as demonstrated in \cite{saracino15} and \cite{massari16}, the PSF variability
due to anisoplanatism effects in MCAO images is too large to be
properly accounted for by a uniform PSF model. The magnitude of the
final MAD catalogue were calibrated as described in \cite{ferraro09}.

The NIRC2 data set consists of 82 exposures in the $K$-filter with
integration time of 180 s, taken during two observing nights (see
Table 1). NIRC2 uses one Laser Guide Star (LGS) to correct for the
blurring effect of the atmosphere. For this reason, the PSF variation
within the field of view is much stronger than what observed in MCAO
data, and the PSF deformation mainly follows a radial pattern around
the LGS itself. In the acquired images, not even a third order
polynomial allowed a proper modeling of the PSF, and a good
photometry was achieved only within the isoplanatic
region. Nonetheless, each exposure has been treated as described for
the previous sets, and a final catalog including only stars found in
at least 5 single exposures was built. The instrumental magnitudes
were calibrated using the MAD catalog as reference.


\section{Unveiling the double Turn-off}
\label{analysis}
The gold standard method to measure the age of a stellar population
requires the accurate definition of the MS-TO region in the
color-magnitude diagram (CMD). In Terzan 5 this is hampered by three
major problems: {\it (i)} strong differential reddening, {\it (ii)}
heavy contamination from disk and bulge field stars and {\it (iii)}
severe stellar crowding.  To overcome these obstacles we took
advantage of previous works published by our group. In particular: the
high-resolution extinction map obtained in the direction of Terzan 5
\citep{massari12} was used to identify regions where the color excess
is the least and the most homogeneous: we excluded from the analysis
all the cluster regions (as the North and the North-East portions of
the system) where the reddening map shows large variations of the
extinction.  Once selected the most appropriate region, we also
applied differential reddening corrections (following the
prescriptions by \citealp{cardelli89}) to the magnitudes of all the
stars detected therein.  To remove possible non-member objects, we
excluded all the sources with relative proper motion larger than 1.5
mas yr$^{-1}$, which likely belong to the Galactic disk and bulge
populations \citep{massari15} \footnote{ All the details about the
    proper motion determination can be found in
    \citet{massari15}. Here we just recall that we used two HST-ACS
    datasets separated by a baseline of $\sim 10$ years, which allowed
    us to measure proper motions with a typical uncertainty of 0.1 mas
    yr${-1}$ at the MS-TO level, thus ensuring good precision in
    assessing the stellar membership.}  In order to remove blends,
only sources with a quite symmetric brightness profile (as measured by
the PSF sharpness parameter) have been considered. Moreover, to
minimize the scatter, only sources within 3-$\sigma$ from the median
photometric error at different magnitude levels, in the $K$-magnitude
versus photometric error diagram, have been taken into
account. Finally, we excluded the innermost and most crowded regions
of the cluster, depending on the field of view and the performances of
each instrument: we excluded stars with $r<16\arcsec$ in the MAD and
NIRC2 samples, those with $r<20\arcsec$ in the ACS sample, and those
with $r<35\arcsec$ in the WFC3 sample.
 
\begin{figure}[b] 
\includegraphics[trim=0cm 0cm 0cm 0cm,clip=true,scale=.80,angle=0]{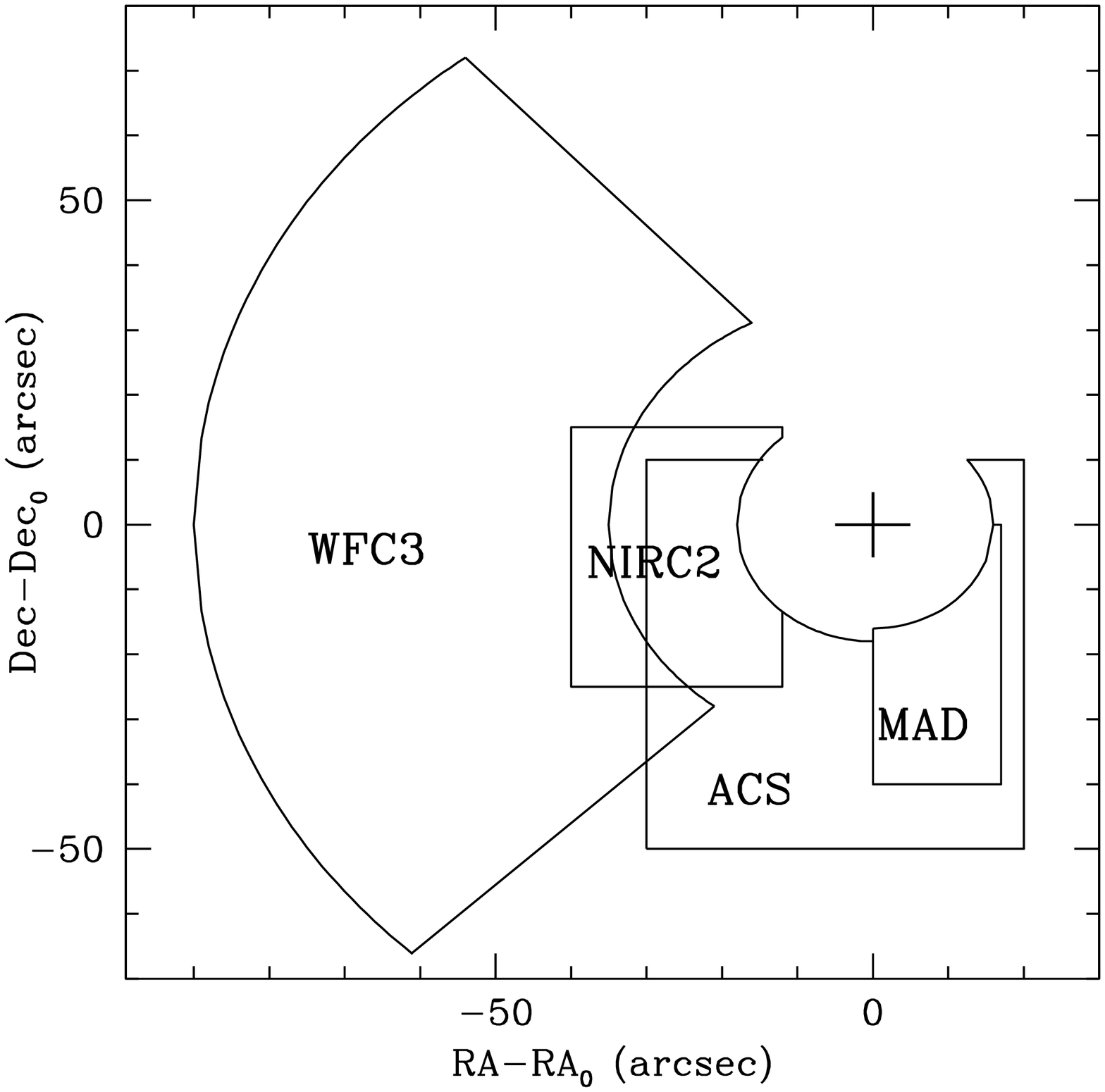} 
\caption{Overview of the regions sampled by the different
  datasets. The plotted regions approximately delimitate the portions
  of Terzan 5 sampled by each dataset and used to construct CMDs in
  different planes (as those shown in Figure 2).  Each portion of the
  dataset is labeled with the name of the camera used to perform the
  observations.}
\label{map} 
\end{figure}

\begin{figure}[b] 
\includegraphics[trim=0cm 0cm 0cm 0cm,clip=true,scale=.80,angle=0]{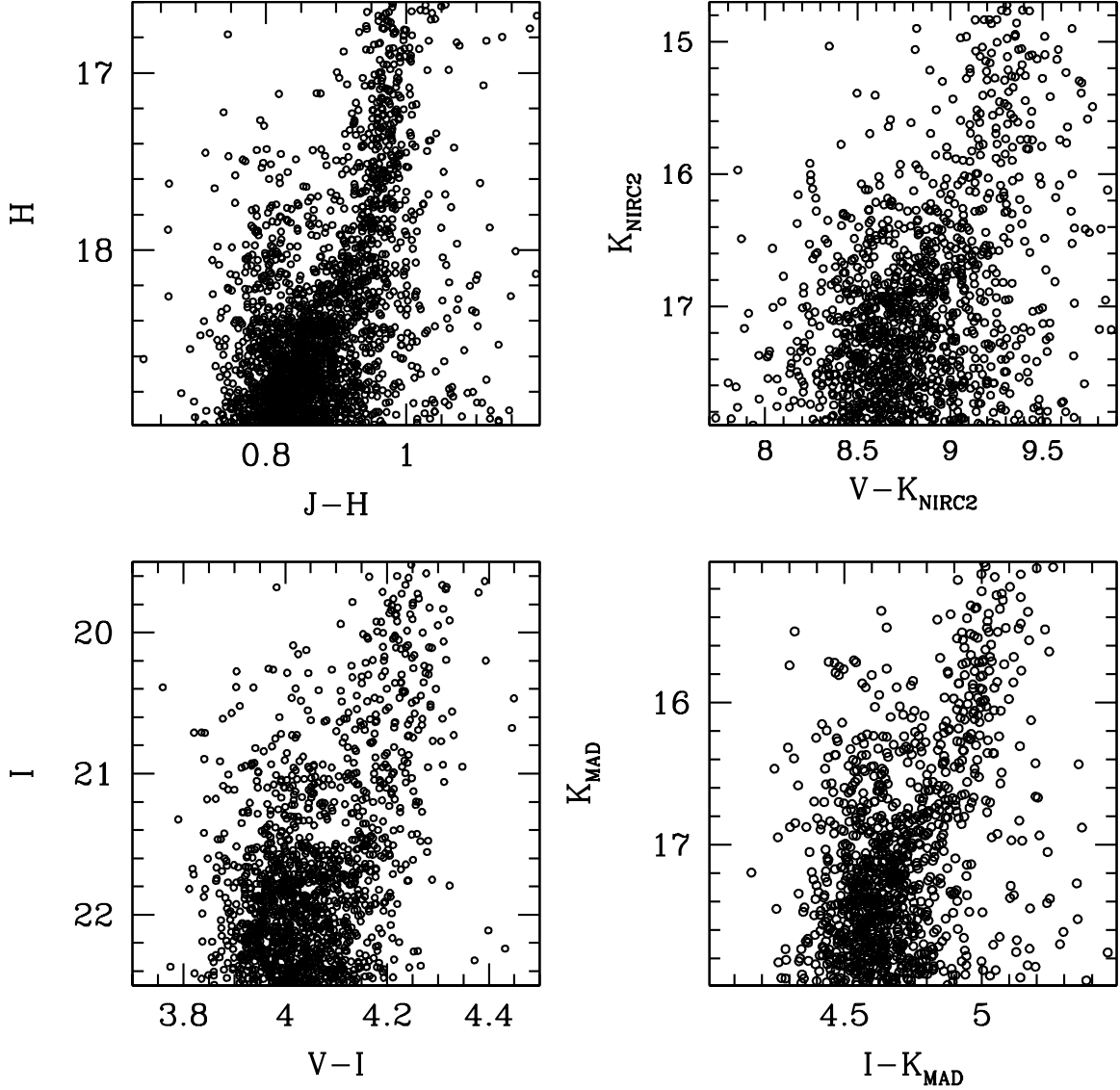} 
\caption{The double MS-To of Terzan 5 in different CMDS. \emph{Upper
    left:} The (H, J-H) CMD obtained from data secured with the
  infrared channel of the HST Wide Field Camera 3 (WFC3).  \emph{Upper
    Right:} The (K, V-K) CMD obtained by combining the K-band laser
  AO-assisted images secured with Keck-NIRC2 and I-band images
  obtained with HST-ACS. \emph{Bottom Left:} The (I, V-I) CMD obtained
  with the HST-ACS.\emph{Bottom Right:} The (K, I-K) CMD obtained by
  combining K-band AO-assisted VLT observations with the deep I-band
  image obtained with HST-ACS. All the CMDs have been corrected for
  internal differential reddening and they refer to different regions
  of the system (see Figure 1). }
\label{cmds} 
\end{figure}

\begin{figure}[b] 
\includegraphics[trim=0cm 0cm 0cm 0cm,clip=true,scale=.80,angle=0]{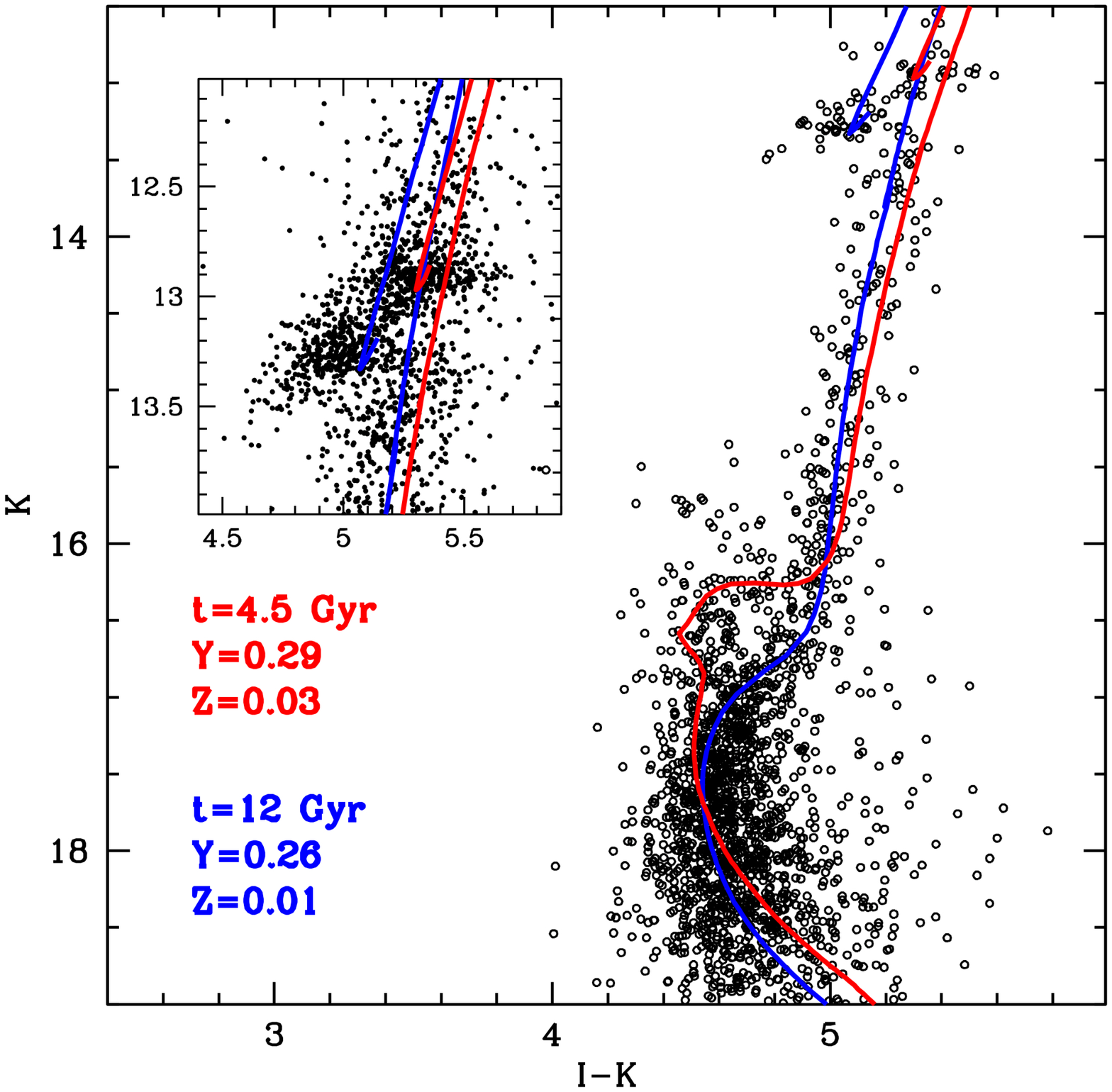} 
\caption{\emph{Top panel:} Differential reddening corrected (K, I-K)
  CMD of Terzan 5 obtained by combining the deepest VLT-MAD K-band and
  HST-ACS I-band images of a relatively small ($\sim600$ arcsec$^2$)
  region, located South-East from the center, where differential
  reddening is the least and the most homogeneous. The CMD clearly
  shows the presence of two distinct MS-TO/SGBs, indicative of two
  stellar sub-populations with different ages. Indeed, in order to
  reproduce both the turnoff regions and the locations of the red
  clumps, two isochrones \citet{girardi02} computed with the observed
  metallicity and assuming significantly different ages are needed:
  the main (sub-solar) population of Terzan 5 is well reproduced
  adopting a mass fraction of heavy elements $Z=0.01$ ([Fe/H]$=-0.3$
  dex), an helium mass fraction $Y=0.26$ and an age $t=12$ Gyr (blue
  line); the super-solar component requires $Z=0.03$ ([Fe/H]$=+0.3$
  dex), $Y=0.29$ and $t=4.5$ Gyr (red line). In order to better
  appreciate the agreement between the isochrones and the data at the
  red clump level, the inset shows a zoomed CMD, with all the measured
  stars plotted (with no selections). }
\label{cmd}
\end{figure}

Figure \ref{map} schematically shows the portions of the cluster field
 covered by each dataset and considered for the analysis. 
 As can be seen, different portions of the South and
South-West regions of Terzan 5 are sampled.  Figure \ref{cmds} shows
the CMD of the TO/SGB region in a few planes, obtained by combining
observations in different photometric bands, under the criteria
described above. Although they refer to different regions of the
cluster, they all show (at different levels of significance) the
existence of two distinct evolutionary sequences at the
MS-TO/sub-giant branch (SGB) level. The clearest one is obtained in
the combined NIR/optical $(K, I-K)$ plane and is plotted in Figure
\ref{cmd}: a well-defined secondary SGB is clearly emerging from the
MS-TO of the dominant population at $K\sim17$, and it merges into the
red giant branch (RGB) at $K\sim16.2$.

\subsection{The age of the two populations}
\label{sec:age} 
\begin{figure}[b] 
\includegraphics[trim=0cm 0cm 0cm 0cm,clip=true,scale=.80,angle=0]{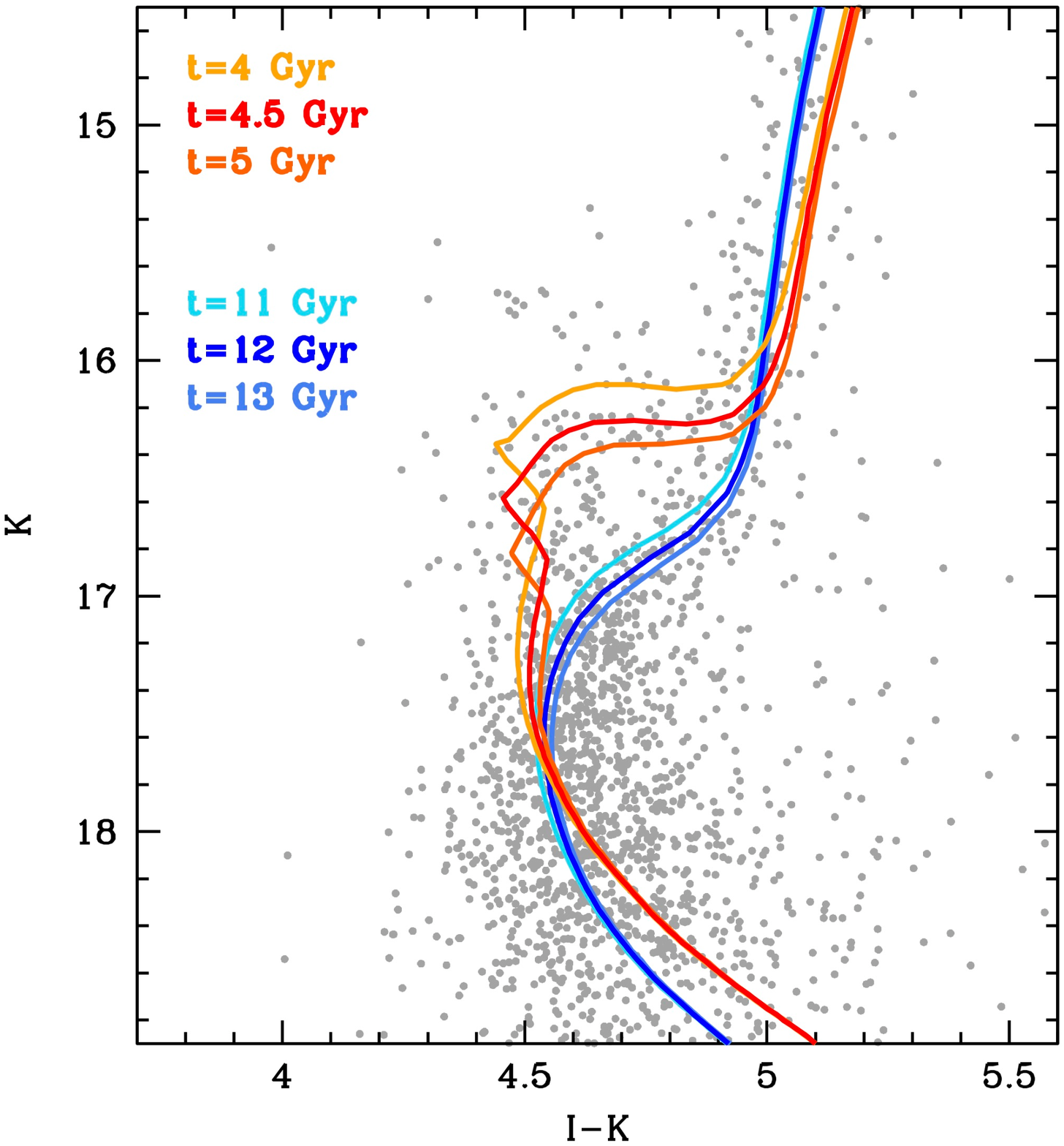} 
\caption{Colour-magnitude diagram of Terzan 5 (same as in Figure 3),
  with isochrones \citep{girardi02} of different ages (see labels)
  superimposed. The three oldest isochrones, computed for Z=0.01 and
  Y=0.26, show that the metal-poor population is consistent with an
  age of $12\pm 1$ Gyr. The three youngest isochrones have $Z=0.03$
  and $Y=0.29$ and show that the metal-rich component of Terzan 5 has
  an age of $4.5 \pm 0.5$ Gyr.}
\label{ages} 
\end{figure}

\begin{figure}[b] 
\includegraphics[trim=0cm 0cm 0cm 0cm,clip=true,scale=.80,angle=0]{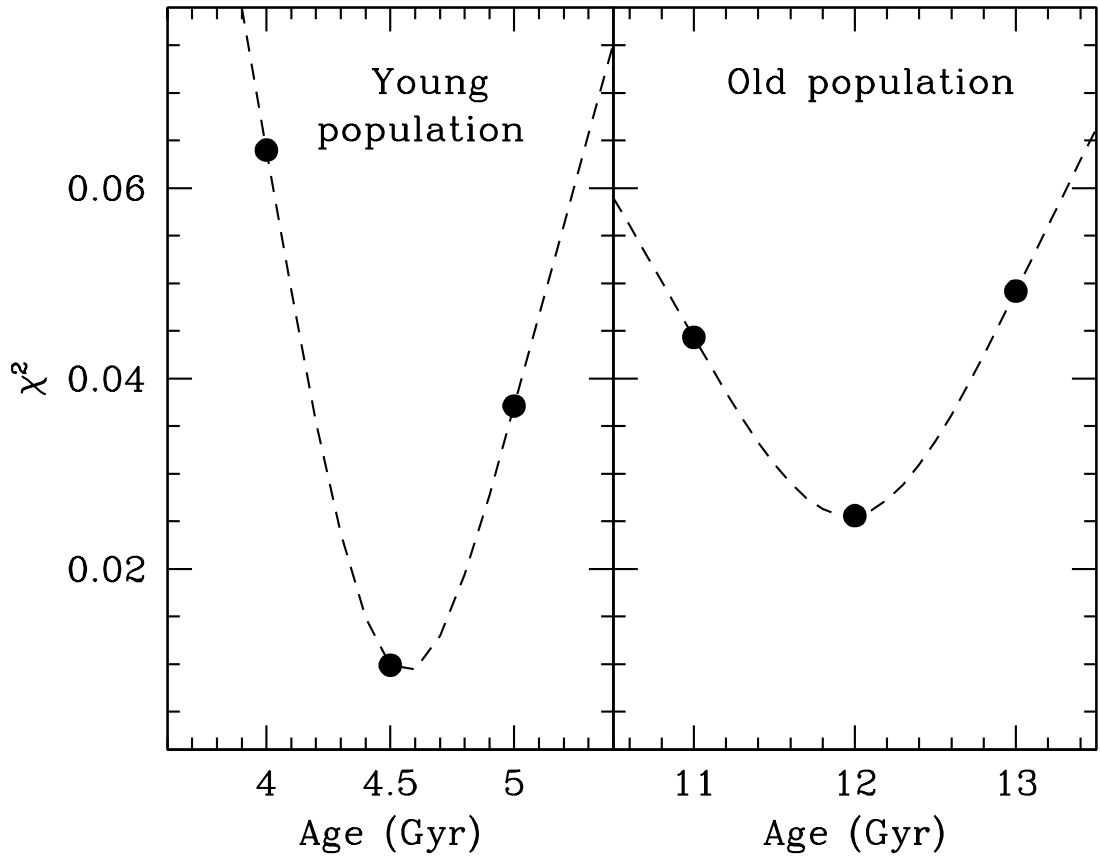} 
\caption{ Isochrone fitting. The value of the $\chi^2$ parameter (see
  text for its definition) is plotted as a function of the age of the
  three isochrones selected to reproduce the young and the old
  populations (see Figure 3). In both cases, a well-defined minimum
  identifies the best-fit isochrone.  }
\label{chi2} 
\end{figure}

In order to determine the age of the two sub-populations hosted in
Terzan 5, we searched for the best-fit isochrones reproducing the
features shown in Figure \ref{cmd}. To this end,  we assumed a
  distance of 5.9 kpc and a color excess $E(B-V)=2.38$
  \citep[from][]{valenti07}, and we adopted the following procedure.
The metallicity of each population ($Z=0.01$ and $Z=0.03$) has been
fixed according to the spectroscopic observations ([Fe/H]$=-0.3$ dex
and [Fe/H]$=+0.3$ dex, respectively), while for the helium mass
fractions, we adopted $Y=0.26$ for the metal-poor component and
$Y=0.29$ for the super-solar one, as expected from ``standard''
$Y-$[Fe/H] enrichment scenarios \citep{girardi02}. Hence, having fixed
the metallicity and the helium mass fraction, the only free parameter
of the fit is the age. In Figure \ref{ages} we show a set of
isochrones with the adopted metallicities and helium mass fractions,
and with different ages selected to match the two detected MS-TO/SGB
features.  In order to identify the isochrones best reproducing the
observations, we performed a Chi-Square analysis. The $\chi^2$
parameter has been computed selecting a sample of stars along the
MS-TO/SGB sequences of the two populations as: $\chi^2 = \Sigma
(O_k-E_k)/E_k$, where $O_k$ is the observed magnitude of each star and
$E_k$ is the corresponding value of the magnitude read along the
isochrone.  The results are shown in Figure \ref{chi2}: the two
distinct MS-TO/SGBs are well reproduced by two isochrones
\citep{girardi02} with quite different ages: $t=12\pm1$ Gyr for the
dominant, sub-solar component, and $t=4.5\pm0.5$ Gyr for the extreme
metal-rich sub-population.

\begin{figure}[b] 
\includegraphics[trim=0cm 0cm 0cm 0cm,clip=true,scale=.80,angle=0]{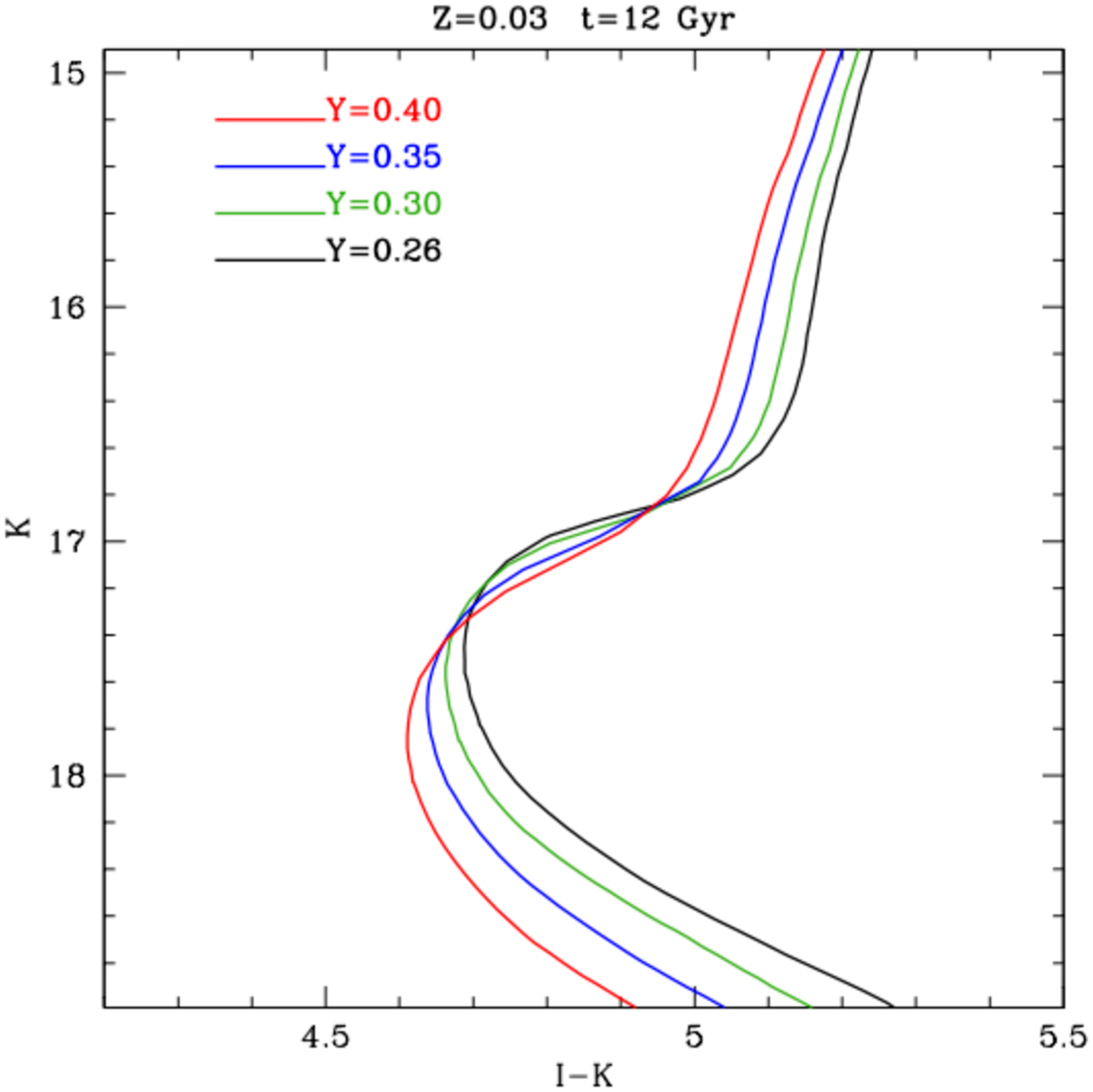} 
\caption{The effect of increasing the helium abundance at fixed age at
  the MS-TO/SGB level.  For illustrative purposes, the MS-TO/SGB
  region of isochrones \citep{girardi02} calculated at fixed age (12
  Gyr) and metallicity (Z=0.03) is shown for increasing helium
  abundances (see labels).  An increased helium abundance makes the
  MS-TO/SGB fainter.}
\label{helium} 
\end{figure}

For sake of completeness, we also investigated the effect of an
increased helium abundance on the isochrones. Indeed in genuine GCs
there is evidence of helium abundances significantly enhanced with
respect to the standard values \citep{piotto07}, and a large
($Y=0.35-0.4$) helium content at fixed old age has been advocated
\citep{dantona10,lee15} to explain the origin of the bright red clump
observed in Terzan 5. Stellar models show that
increasing the helium abundance at fixed age makes the MS-TO/SGB
fainter (see Figure \ref{helium}), at variance  to what is
needed to reproduce the secondary MS-TO observed in Terzan 5.
The best fit isochrones (of 12 and 4.5 Gyr and standard
helium) nicely reproduce also the location of the two red clumps, thus
demonstrating that even this feature is likely due to an age difference,
instead of a different helium abundance in two almost coeval
populations.  Hence the detection of the bright TO completely
removes the possibility that the two stellar populations are due to a
difference in helium and definitely breaks the helium/age degeneracy.

\subsection{The radial distribution of the two populations}
\label{radist}
\begin{figure}[b] 
\includegraphics[trim=0cm 0cm 0cm 0cm,clip=true,scale=.85,angle=0]{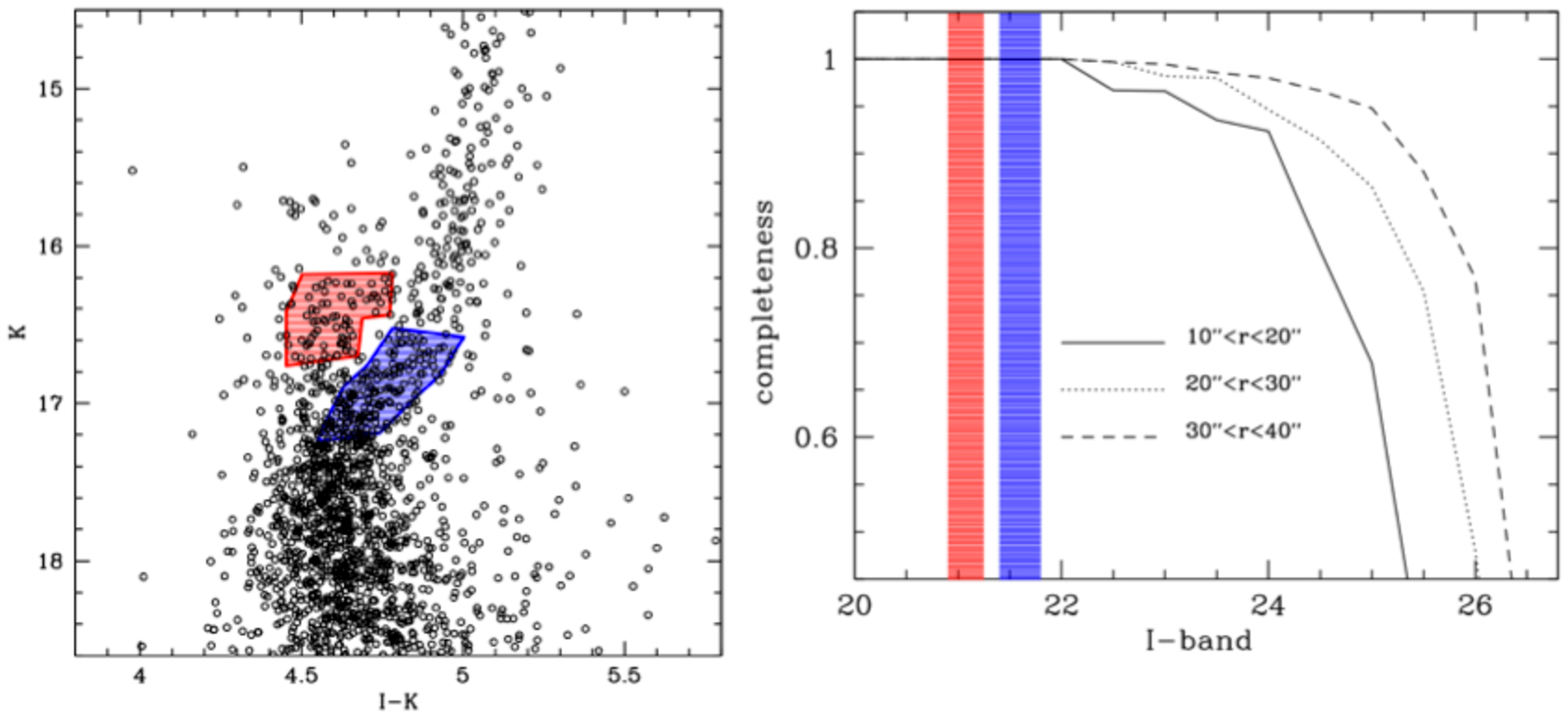} 
\caption{Population selection and completeness analysis for the
  cumulative radial distribution study. \emph{Top panel:} The boxes
  adopted to select stars along the MS-TO/SGB sequences of the old
  (blue shaded region) and young (red shaded region) populations. All
  the stars with $r>10\arcsec$ lying within the two selection boxes
  have been considered to construct the radial distributions shown in
  the upper panel of Figure 8. \emph{Bottom panel:} The completeness
  curves in the I-band obtained in radial annuli at different distances
  from the centre (see labels). The two vertical strips mark the
  I-band magnitude ranges ($20.9<I<21.35$ in red, and $21.4<I<21.8$ in
  blue) covered by the boxes adopted to select, respectively, the
  young and the old populations used to study the cumulative radial
  distributions.}
\label{comple}
\end{figure}

The super-solar red clump sub-population
was found   to be more centrally
concentrated than the more metal poor one\citep{ferraro09,lanzoni10}.  
In order to study the
radial distribution of the two MS-TO/SGB populations, we first
considered the CMD shown in Figure \ref{cmd}, where the two sequences
are clearly distinguishable, and adopted the sample selection boxes
marked as colored shaded regions in the left-hand panel of Figure
\ref{comple}. Then, all the stars measured in the $K$ and $I$ bands
and located within the two selection boxes have been considered, with
the only exception (for avoiding incompleteness biases) of the 
innermost stars within $10\arcsec$ from the center.  To investigate the
completeness of the samples thus selected we followed the standard
procedures of adding artificial stars to the original frames. The
results obtained for the $I$-band (providing the main source of
incompleteness) are shown in the right-hand panel of Figure
\ref{comple}, where the two shaded strips mark the $I$-band magnitude
range of the young ($20.9<I<21.35$, in red) and of the old
($21.4<I<21.8$ in blue) population selection boxes.  As can be seen,
at these magnitude levels the completeness of both the samples is
$100\%$.

\begin{figure}[b] 
\includegraphics[trim=0cm 0cm 0cm 0cm,clip=true,scale=.80,angle=0]{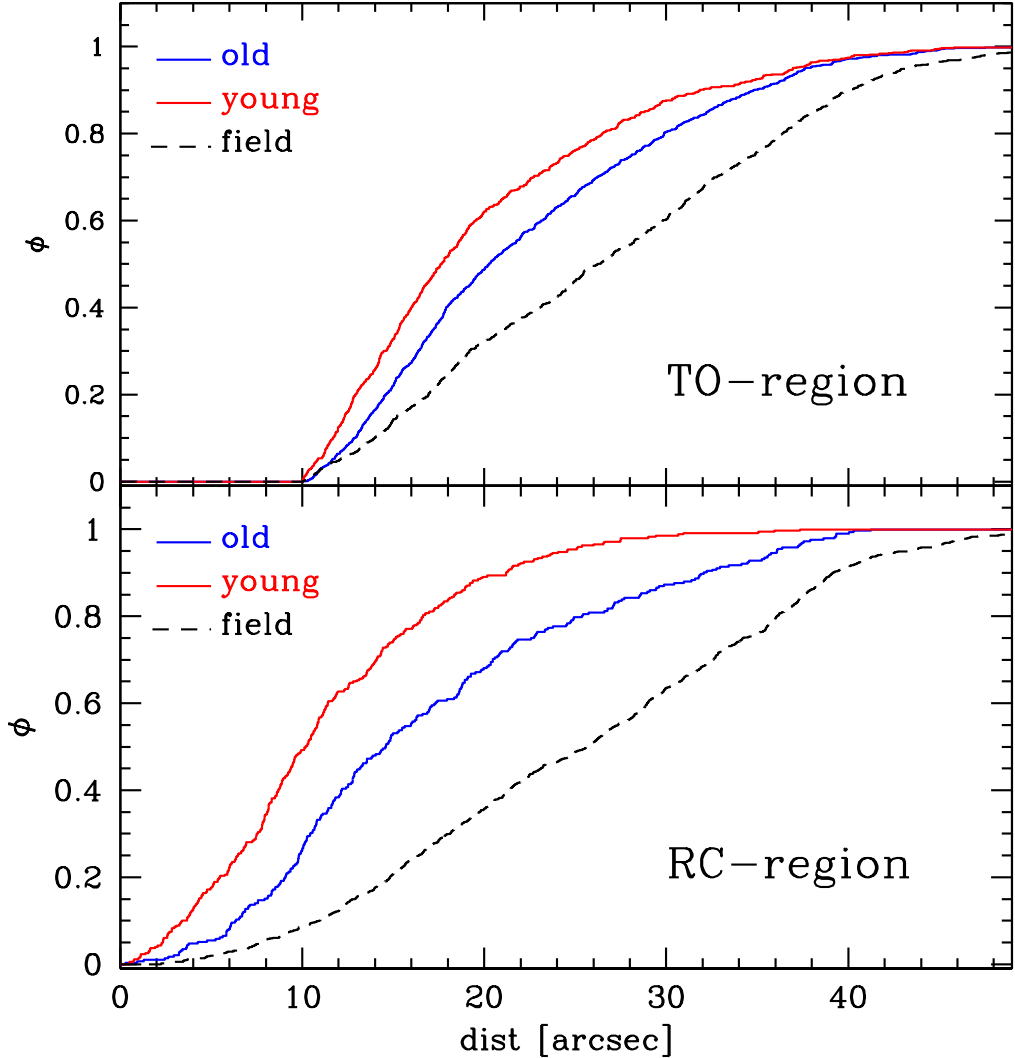} 
\caption{\emph{Top panel:} Cumulative radial distribution of MS-TO/SGB
  stars belonging to the young population (red line) and to the old
  population (blue line), compared to that of field stars (black
  dashed line), as a function of the projected distance from the
  center of Terzan 5. The samples of MS-TO/SGB stars have been
  selected on the basis of the $(K, I-K)$ CMD as shown in Figure 7. In
  order to avoid incompleteness biases, only stars at $r>10\arcsec$
  from the center have been considered.  The field distribution has
  been obtained from a synthetic sample of 10,000 stars uniformly
  distributed over the considered observed region.  \emph{Bottom
    Panel:} The same as in the top panel, but for stars selected from
  the two red clumps with no radial selection.  Clearly, the young
  population is more centrally concentrated than the old one. }
\label{ks}
\end{figure}

The final samples count about 670 stars along the young MS-TO/SGB and
900 stars along the old MS-TO/SGB.  The radial distribution of the two
samples is shown in Figure \ref{ks} and demonstrates that the young
component is clearly more centrally concentrated than the old one.
This finding is fully consistent with what found from the analysis of
the double clumps \citep{ferraro09, lanzoni10} and in agreement with
what expected in the context of a self-enrichment scenario, where the
younger population formed from a central concentration of gas enriched
from the previous stellar generation.

In order to investigate whether Galactic field contamination
significantly affects the selected populations, we estimated the field
star radial distribution. Indeed, Galactic field stars are expected to
be uniformly distributed over the small region covered by the
 observations, while any stellar populations belonging to
Terzan 5 must be concentrated toward the center of the system. Hence,
the field star radial distribution has been derived from a simulation
of 10,000 artificial stars with a uniform spatial distribution in the
observed field of view. The field distribution (black dashed lines in
Figure 8) turns out to be totally incompatible with that of both the
Terzan 5 sub-populations.


\section{Discussion and Conclusions}
\label{discussion}
Both the multi-peak iron distribution \citep{ferraro09, origlia11,
  origlia13, massari14} and the clear separation of the two MS-TO/SGBs
(this paper) observed in Terzan 5 suggest that the star formation
process in the system was not continuous, but characterized by
distinct bursts. Dating the two main sub-populations provides the time
scale of the enrichment process.  After an initial period
of star formation, which occurred at the epoch of the original
assembly of the system ($\sim 12$ Gyr ago) and generated the two
sub-solar components from gas enriched by SNII, Terzan 5 experienced a
long phase ($t \sim 7.5$ Gyr) of quiescence during which the gas
ejected from both SNeII and SNeIa accumulated in the central region.
Then, approximately 4.5 Gyr ago, the super solar component
formed. This requires the Terzan 5 ancestor to be quite massive (at
least a few $10^8 M_\odot$; \citealt{baum08}).  The amount of gas
retained by the proto-Terzan 5 system was huge: on the basis of the
observed red clump populations \citep{ferraro09,lanzoni10}, we
estimate that the young component in Terzan 5 currently amounts to
roughly $7.5\times 10^5 M_\odot$ (the same size of a 47 Tucanae-like
globular cluster) corresponding approximately to $38\%$ of the total
mass of the system.

These results definitely put Terzan 5 outside the context of genuine
CGs. Indeed, apart from the overall morphological appearance, Terzan 5
shares no other property with globulars, neither in terms of chemistry
(GCs show inhomogeneities only in the light-elements;
\citealt{carretta09}), nor in the enrichment history and age spread of
the sub-populations (the enrichment timescale in GCs is of a few
$10^8$ yrs and their light-element sub-populations are thus almost
coeval; \citealt{dercole08}), nor in terms of the mass of the
progenitor (GCs did not retain the high-velocity SN ejecta and
therefore do not need to have been as massive as Terzan 5 in the past;
\citealt{baum08}).  {\it If Terzan5 is not a genuine GC, what is it
  then?}

\begin{figure}[b] 
\includegraphics[trim=0cm 0cm 0cm 0cm,clip=true,scale=.80,angle=0]{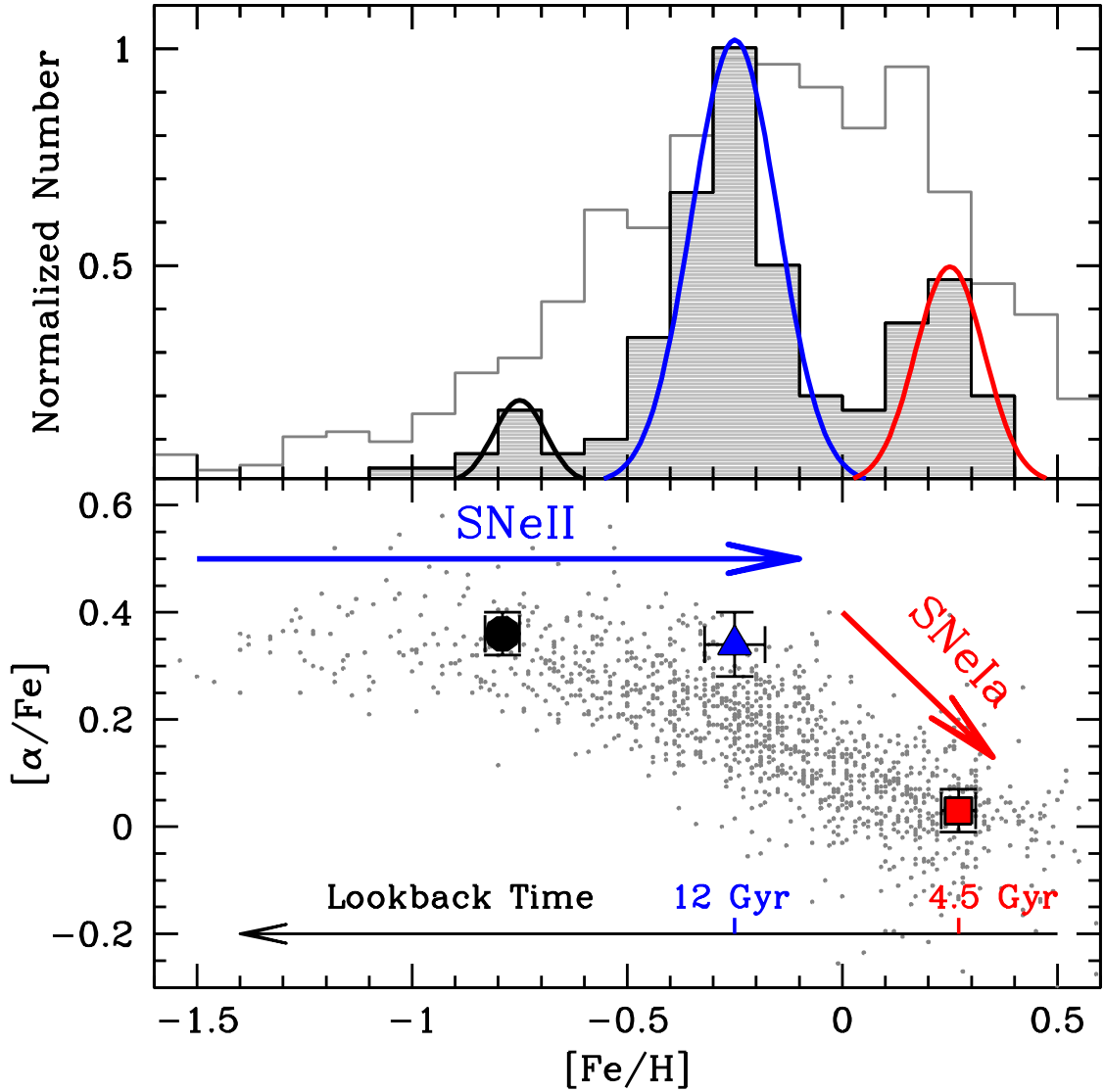} 
\caption{\emph{Top panel:} Iron distribution of the three
  sub-populations of Terzan 5 (grey histogram and colored lines)
  compared to that of the Galactic bulge field stars from the
  literature \citep{ness13,johnson14}.  \emph{Bottom Panel:}
  [$\alpha$/Fe]-[Fe/H] distribution of the three sub-populations of
  Terzan 5 (extreme metal-poor: black circle; sub-solar: blue
  triangle; super-solar: red square) compared to that of the bulge
  field stars from the literature (grey dots). The formation epoch of
  the two major Terzan 5 sub-populations is also labeled.}
\label{alpha} 
\end{figure}

As shown in Figure \ref{alpha}, the metallicity distribution and
chemical abundance patterns of Terzan 5 are strikingly similar to
those observed in bulge stars
\citep{gonzalez11,ness13,johnson14,rojas14,ryde16}: not only the iron
abundance, but also the amount of $\alpha$-enhancement and its
dependence on metallicity are fully consistent with those measured in
the Galactic bulge. Note that setting such an abundance pattern
requires quite special conditions. In fact, the large value of [Fe/H]
($\sim -0.2$) at which the $[\alpha/$Fe] vs [Fe/H] relation changes
slope necessarily implies that an exceptionally large number of SNeII
exploded over a quite short timescale (i.e. before the explosion of
the bulk of SNeIa, which polluted the medium with iron, generating the
knee in the relation). In turn, this requires a very high star
formation efficiency. Indeed, the stellar populations in the Galactic
halo/disk system \citep{chiappini99,matteucci05} and in the dwarf
galaxies observed in the vicinity of the Milky Way \citep{lemasle12}
show abundance patterns inconsistent with those plotted in the figure:
they do not reach the same high iron content and the knee in the
$[\alpha/$Fe] vs [Fe/H] relation occurs at much lower metallicities
([Fe/H]$\sim -1$ and [Fe/H]$\sim -1.5$ in the disk/halo and in dwarfs,
respectively), indicating a significantly less efficient SNII
enrichment process.  Hence, within the Local Group, the abundance
patterns shown in Figure \ref{alpha} can be considered as a univocal
signature of  stellar populations toward the Galactic bulge.
 
Because of the tight chemical link between Terzan 5 and the bulge, it
seems reasonable to ask whether there is any connection also in terms
of stellar ages, in particular, between the super-solar (and young)
population of Terzan 5 and the super-solar component of the bulge.
Deep photometric studies \citep{clarkson08,valenti13} of a few bulge
fields properly decontaminated from the disk population indicate that
the vast majority of the bulge is significantly old (up to $80\%$ is
older than 10 Gyr; \citealt{clarkson08, nataf15}).  However, a
multi-peak metallicity distribution very similar to that of Terzan 5
is observed in samples of giant, red clump, and lensed dwarf stars in
the bulge (\citealp{ness13, rojas14, bensby13}, respectively; but see
\citealp{johnson14}). Moreover, in close analogy with what found for
Terzan 5, two prominent epochs of star formation are estimated for the
lensed dwarfs (more than 10 Gyr ago for stars at [Fe/H]$<0$, and $\sim
3$ Gyr ago for the youngest super-solar objects;
\citealp{bensby13,nataf15}).
 
The possible presence of a younger population was also recently
  suggested in the photometric and spectroscopic study of the bulge
  within the GAIA-ESO Survey.  \citet{rojas14} detected a bimodal
  magnitude distribution for the two RC samples with different
  metallicities observed in their surveyed field closest to the
  Galactic plane (the Baade's Window), where geometric effects due to
  the X-shaped morphology of the bulge should be negligible, thus
  concluding that this might be due to an intrinsic age difference of
  the order of 5 Gyr. Interestingly enough, the difference in
  magnitude ($\Delta K \sim 0.3$), the metallicities (super-solar in
  the bright RC and sub-solar in the fainter one) and the proposed
  difference in age ($\Delta t \sim 5$ Gyr) of their two RC
  populations turn out to be in nice agreement with those measured in
  the two main sub-populations of Terzan5.

The collected evidence suggests that the proto-Terzan 5 system was a
massive structure formed at the epoch of the Milky Way bulge
formation. Indeed, $10^8-10^9 M_\odot$ clumps are observed in early
disks at high redshifts, thus confirming that such massive structures
existed in the remote epoch of the Galaxy assembling ($\sim 12$ Gyr
ago), when also Terzan 5 formed.  These giant clumps are thought to
form from the fragmentation of highly unstable disks, to experience
intense star formation episodes over quite short timescales (less than
1 Gyr)\footnote{The possibility of such clumps to survive the
  energetic feedback from young stars has been critically discussed in
  the literature \citep[see, e.g.,][]{hopkins12}. However recent
  observations suggest that indeed these structure can survive for
  several $10^8$ yr (a lifetime of 500 Myr has been estimated by
  \citet{zanella15}, thus supporting the scenario where these clumps
  can contribute to bulge formation.} and then to coalesce and give
rise to galaxy bulges (e.g., \citealp{elmegreen08}).  Notably the
metallicity and the $\alpha$-element enhancement of the high-z clumps
([Fe/H]$= -0.6$ and $[\alpha/$Fe$]\sim 0.25-0.7$;
\citealp{pettini02,halliday08}) are in agreement with the values
measured in the two oldest stellar populations of Terzan 5 (and in the
Galactic bulge), thus suggesting that the major star formation events
occurred at early epochs and over a quite short time-scale, before the
bulk of SNeIa exploded.  Hence the progenitor of Terzan 5 might have
been one of these clumps, which survived the complete destruction.
Most of the pristine gas was frenetically converted into stars within
the giant clumps that populated the forming bulge.  The coalescence of
most of these systems in the very early epoch  contributed to
  form the ``old bulge'' that we currently observe.\footnote{The
  proposed scenario might also naturally account for the $\gamma$-ray
  emission recently detected in the central region of our Galaxy
  \citep{brandt15}. This has been proposed to be originated by
  Millisecond Pulsars dispersed in the field by the dissolution of
  globular clusters in the early epoch of the Galaxy
  formation. Further support to this hypothesis is obtained by
  assuming that the disrupting structures were metal-rich and massive
  clumps similar to the proposed progenitor of Terzan 5.  In fact, a
  high metallicity (testifying the explosion of a large number of
  SNeII; see Figure 9) combined with a large total mass (allowing the
  retention of both SNeII ejecta and neutron stars), and with a high
  collision rate (as measured by \citet{lanzoni10} for Terzan 5),
  makes the proto-bulge massive clumps the ideal environment for
  forming Millisecond Pulsars. Indeed, Terzan 5 is the stellar system
  hosting the largest known population of such objects in the Galaxy
  \citep{ransom05}.  The large number of Millisecond Pulsars generated
  within the clumps and then dispersed in the bulge may therefore
  account for the $\gamma$-ray emission detected in the Galactic
  center. Note that, at odds with pulsars, the emission of the
  reaccelerated pulsars is expected \citep{bhatta91} to persist for
  long time ($10^{10}$ yr).}  However, a few of them survived the
destruction.  A possibility is that the proto-Terzan5 system was
originally less massive \citep{behrendt16} and more compact than other
clumps and, because of its lower mass, it did not rapidly sink and
merge in the central region, but it was scattered out to large radii
by dynamical interactions with more massive sub-structures (indeed
such events are observed to occur in simulations;
\citealt{bournaud16}). Thus the proto-Terzan5 clump might have evolved
in an environment significantly less violent than the central forming
bulge, and, as a giant ``cocoon'', it may have retained a large amount
of gas ejected by SNe for several Gyr, before producing a new
generation of stars (such a bursty star formation histories, with long
periods of quiescence, are observed in the Universe; e.g.,
\citealp{van15}). For instance, the most recent burst in Terzan 5
could have been promoted by a major interaction with bulge
sub-structures, possibly also driving the mass-loss phenomenon that
significantly reduced its mass (from the initial $\sim 10^8 M_\odot$,
to the current $10^6 M_\odot$; \citealt{lanzoni10}).
 
An alternative possibility is that Terzan 5 is the former nuclear
cluster of a massive satellite that contributed to generate the Milky
Way bulge via repeated mergers at high redshift
\citep[e.g.][]{hopkins10}.  In order to account for the observed
chemical abundance patterns (Figure \ref{alpha}), the satellite
hosting the Terzan 5 progenitor should have been much more metal rich
(and massive) than the dwarf galaxies currently populating the Local
Group.  Although such objects are not present anymore in the Milky Way
vicinity \citep{chiappini99,matteucci05,lemasle12}, they possibly
existed at the epoch of the Galaxy assembly\footnote{On the other
  hand, a scenario where the structure hosting the proto-Terzan 5 was
  accreted in a single event well after the bulge formation seems to
  be quite implausible, since it would require that such a structure
  independently experienced a chemical evolutionary history very
  similar to that of the bulge (see Figure \ref{alpha}).}.  Thus, the
proto-Terzan 5 host might have been one of several satellites that
merged together to originate the bulge \citep{hopkins10}.  A
speculation on its mass can be derived from the mass-metallicity
relation of $z=3-4$ galaxies \citep{mannucci09}, by assuming that the
satellite was somehow ($<0.5$ dex) more metal-poor than its nucleus
(which seems to be the case at least for the most massive structures;
see Figure 4 in \citealt{paudel11}). This yields to a mass of the
proto-Terzan5 host galaxy of a few $10^9 M_\odot$.   Thus similar
  primordial massive structures (see \citealt{hopkins10}) could have
  contributed to the mass budget of the Galactic bulge ($2 \times
  10^{10} M_\odot$; \citealt{valenti16}).


The bulge formation history in general, its broad metallicity
  distribution and its kinematics cannot be explained within a simple
  scenario (either merging, dissipative collapse or some disk
  instability/secular evolution), but likely all these processes and
  perhaps others have played some role.  Complexity is a matter of
  fact in all the galactic components, especially in the central
  region of the Galaxy. Any ad hoc model that neglects complexity to
  support one specific scenario is poorly thorough. While the presence
  of bar(s) and disks with complex chemistry and kinematics in the
  central region of our Galaxy is well established since quite a long
  time, an observational evidence of the existence of fossil remnants
  of massive clumps in our bulge, as those observed at high-z, is
  still lacking.  Terzan 5 could be the first such evidence, providing
  a direct link between the distant and the local Universe and also
  calling for models able to account for the survival of similar
  structures and for the existence of a younger (minor) stellar
  populations in galaxy bulges.  From an observational point of view,
  it is urged to check whether other stellar systems similar to Terzan
  5 still orbit the Milky Way bulge and to constrain the amount of
  intermediate-age stars therein.  Finally, it is also worth
  mentioning that the presence of massive clumps that merged/dissolved
  to form the bulge, or survived disruption and evolved as independent
  systems, does not necessarily exclude the formation (and possible
  secular evolution) of disks, bars and other sub-structures as those
  observed toward the Galactic bulge.

\acknowledgements This research is part of the project {\it
  Cosmic-Lab} (see http://www.cosmic-lab.eu) funded by the European
Research Council (under contract ERC-2010-AdG-267675).


\begin{deluxetable}{lrcccll}
\tablecolumns{7}
\tiny
\tablewidth{0pt}
\tablecaption{Observational Datasets}
\tablehead{\colhead{Instrument} & \colhead{N$_{\rm exp}$} & \colhead{$t_{\rm exp}$}
& \colhead{Filters} & \colhead{Epoch} & \colhead{Proposal} & \colhead{PI} \\
 \colhead{Telescope}&   & \colhead{(sec)} & & & & } 
\startdata
 & & & & & &   \\
 MAD@ESO-VLT   &  15   & 120   & J,K   & 2008    & Science Dem.  & FERRARO         \\
 NIRC2@KECK   &  29   & 180   & K   & 2010    & U156N2L   & RICH         \\
 WFC3@HST   &  64   & 300   & F110W, F160W   &  2013    & GO12933   & FERRARO      \\ 
ACS@HST   &  10   & 365   & F606W, F814W   &  2013    & GO12933   & FERRARO       \\
ACS@HST   &  2   & 340   & F606W, F814W   &  2003    & GO9799   & RICH          \\
ACS@HST   &  10   & 365   & F606W, F814W   & 2015    & GO14061   & FERRARO           \\
\enddata
\label{tab1}
\end{deluxetable}


\begin{thebibliography}
  
\bibitem[Anderson \& Bedin(2010)]{jaybedin10} Anderson, J., \& Bedin,
  L.~R.\ 2010, \pasp, 122, 1035

\bibitem[Ballero et al.(2007)]{ballero07} Ballero, S.~K., Matteucci,
  F., Origlia, L., \& Rich, R.~M.\ 2007, \aap, 467, 123

\bibitem[Baumgardt et al.(2008)]{baum08} Baumgardt, H., Kroupa, P., \&
  Parmentier, G.\ 2008, \mnras, 384, 1231

\bibitem[Bhattacharya \& van den Heuvel(1991)]{bhatta91} Bhattacharya,
  D., \& van den Heuvel, E.~P.~J.\ 1991, \physrep, 203, 1

\bibitem[Behrendt et al.(2016)]{behrendt16} Behrendt, M., Burkert, A.,
  \& Schartmann, M.\ 2016, \apjl, 819, L2

\bibitem[Bensby et al.(2013)]{bensby13} Bensby, T., Yee, J.~C.,
  Feltzing, S., et al.\ 2013, \aap, 549, A147

\bibitem[Bournaud(2016)]{bournaud16} Bournaud, F.\ 2016, Galactic
  Bulges, 418, 355

\bibitem[Brandt \& Kocsis(2015)]{brandt15} Brandt, T.~D., \& Kocsis,
  B.\ 2015, \apj, 812, 15

\bibitem[Cardelli et al.(1989)]{cardelli89} Cardelli, J.~A., Clayton,
  G.~C., \& Mathis, J.~S.\ 1989, \apj, 345, 245

\bibitem[Carollo et al.(2007)]{carollo07} Carollo, C.~M., Scarlata,
  C., Stiavelli, M., Wyse, R.~F.~G., \& Mayer, L.\ 2007, \apj, 658,
  960

\bibitem[Carretta et al.(2009)]{carretta09} Carretta, E., Bragaglia,
  A., Gratton, R., \& Lucatello, S.\ 2009, \aap, 505, 139

\bibitem[Chiappini et al.(1999)]{chiappini99} Chiappini, C.,
  Matteucci, F., Beers, T.~C., \& Nomoto, K.\ 1999, \apj, 515, 226

\bibitem[Clarkson et al.(2008)]{clarkson08} Clarkson, W., Sahu, K.,
  Anderson, J., et al.\ 2008, \apj, 684, 1110-1142

\bibitem[Combes \& Sanders(1981)]{combes81} Combes, F., \& Sanders,
  R.~H.\ 1981, \aap, 96, 164

\bibitem[Dalessandro et al.(2015)]{dalessandro15} Dalessandro, E.,
  Ferraro, F.~R., Massari, D., et al.\ 2015, \apj, 810, 40

\bibitem[D{\'e}k{\'a}ny et al.(2015)]{dekany15} D{\'e}k{\'a}ny, I.,
  Minniti, D., Majaess, D., et al.\ 2015, \apjl, 812, L29

\bibitem[D'Antona et al.(2010)]{dantona10} D'Antona, F., Ventura, P.,
  Caloi, V., D'Ercole, A., Vesperini, E., Carini, R., \& Di
  Criscienzo, M., 2010, ApJ, 715, L63

\bibitem[D'Ercole et al.(2008)]{dercole08} D'Ercole, A., Vesperini,
  E., D'Antona, F., McMillan, S.~L.~W., \& Recchi, S.\ 2008, \mnras,
  391, 825
 
\bibitem[Elmegreen et al.(2008)]{elmegreen08} Elmegreen,
  B.~G.,Bournaud, F., \& Elmegreen, D.~M.\ 2008, \apj, 688, 67

 
\bibitem[Ferraro et al.(2009)]{ferraro09} Ferraro, F.~R., Dalessandro,
  E., Mucciarelli, A., et al.\ 2009, \nat, 462, 483

\bibitem[Ferraro et al.(2015)]{ferraro15} Ferraro, F.~R., Pallanca,
  C., Lanzoni, B., et al.\ 2015, \apjl, 807, L1
 
\bibitem[Girardi et al.(2002)]{girardi02} Girardi, L., Bertelli, G.,
  Bressan, A., et al.\ 2002, \aap, 391, 195

\bibitem[Grieco et al.(2012)]{grieco12} Grieco, V., Matteucci, F.,
  Pipino, A., \& Cescutti, G.\ 2012, \aap, 548, A60

\bibitem[Gonzalez et al.(2011)]{gonzalez11} Gonzalez, O.~A., 
Rejkuba, M., Zoccali, M., et al.\ 2011, \aap, 530, A54 
 
\bibitem[Halliday et al.(2008)]{halliday08} Halliday, C., Daddi, E.,
  Cimatti, A., et al.\ 2008, \aap, 479, 417

\bibitem[Hopkins et al.(2010)]{hopkins10} Hopkins, P.~F., Bundy, K.,
  Croton, D., et al.\ 2010, \apj, 715, 202

\bibitem[Hopkins et al.(2012)]{hopkins12} Hopkins, P.~F., Kere{\v s},
  D., Murray, N., Quataert, E., \& Hernquist, L.\ 2012, \mnras, 427,
  968

\bibitem[Immeli et al.(2004)]{immeli04} Immeli, A., Samland, M.,
  Gerhard, O., \& Westera, P.\ 2004, \aap, 413, 547
 
\bibitem[Johnson et al.(2014)]{johnson14} Johnson, C.~I., Rich, R.~M.,
  Kobayashi, C., Kunder, A., \& Koch, A.\ 2014, \aj, 148, 67

\bibitem[Lanzoni et al.(2010)]{lanzoni10} Lanzoni, B., Ferraro, F.~R.,
  Dalessandro, E., et al.\ 2010, \apj, 717, 653
 
\bibitem[Lee et al.(2015)]{lee15} Lee, Y.-W., Joo, S.-J., \& Chung,
  C.\ 2015, \mnras, 453, 3906

\bibitem[Lemasle et al.(2012)]{lemasle12} Lemasle, B., Hill, V.,
  Tolstoy, E., et al.\ 2012, \aap, 538, A100

\bibitem[Mannucci et al.(2009)]{mannucci09} Mannucci, F., Cresci, G.,
  Maiolino, R., et al.\ 2009, \mnras, 398, 1915

\bibitem[Massari et al.(2012)]{massari12} Massari, D., Mucciarelli,
  A., Dalessandro, E., et al.\ 2012, \apjl, 755, L32

\bibitem[Massari et al.(2014)]{massari14} Massari, D., Mucciarelli,
  A., Ferraro, F.~R., et al.\ 2014, \apj, 795, 22
  
\bibitem[Massari et al.(2015)]{massari15} Massari, D., Dalessandro,
  E., Ferraro, F.~R., et al.\ 2015, \apj, 810, 69

\bibitem[Massari et al.(2016)]{massari16} Massari, D., Fiorentino, G.,
  McConnachie, A., et al.\ 2016, \aap, 586, A51

\bibitem[Matteucci \& Chiappini(2005)]{matteucci05} Matteucci, F., \&
  Chiappini, C.\ 2005, \pasa, 22, 49

 
\bibitem[McWilliam et al.(2008)]{mcwilliam08} McWilliam, A.,
  Matteucci, F., Ballero, S., et al.\ 2008, \aj, 136, 367

 
\bibitem[Ness et al.(2013)]{ness13} Ness, M., Freeman, K.,
  Athanassoula, E., et al.\ 2013, \mnras, 430, 836

\bibitem[Nataf(2015)]{nataf15} Nataf, D.~M.\ 2015, In 
Fifty Years of Wide Field Studies in the Southern Hemisphere: Resolved Stellar
Populations of the Galactic Bulge and Magellanic Clouds. ASP Conference Series, Vol. 491, 
(San Francisco: Astronomical Society of the
Pacific),p.174
 
 
\bibitem[Origlia(2014)]{origlia14} Origlia, L.\ 2014, 
Setting the scene for Gaia and LAMOST, IAU Symposium Vol 298, 28 
 
 
\bibitem[Origlia et al.(2011)]{origlia11} Origlia, L., Rich, R.~M.,
  Ferraro, F.~R., et al.\ 2011, \apjl, 726, L20

\bibitem[Origlia et al.(2013)]{origlia13} Origlia, L., Massari, D.,
  Rich, R.~M., et al.\ 2013, ApJL, 779, L5 

\bibitem[Paudel et al.(2011)]{paudel11} Paudel, S., Lisker, T., \&
  Kuntschner, H.\ 2011, \mnras, 413, 1764

\bibitem[Pettini et al.(2002)]{pettini02} Pettini, M., Rix, S.~A.,
  Steidel, C.~C., et al.\ 2002, \apj, 569, 742

\bibitem[Piotto et al.(2007)]{piotto07} Piotto, G., Bedin, L.~R.,
  Anderson, J., et al.\ 2007, \apjl, 661, L53

\bibitem[Raha et al.(1991)]{raha91} Raha, N., Sellwood,
  J.~A., James, R.~A., \& Kahn, F.~D.\ 1991, \nat, 352, 411

\bibitem[Ransom et al.(2005)]{ransom05} Ransom, S.~M., Hessels,
  J.~W.~T., Stairs, I.~H., et al.\ 2005, Science, 307, 892

\bibitem[Saha \& Gerhard(2013)]{saha13} Saha, K., \& Gerhard,
  O.\ 2013, \mnras, 430, 2039

\bibitem[Saracino et al.(2015)]{saracino15} Saracino, S., Dalessandro,
  E., Ferraro, F.~R., et al.\ 2015, \apj, 806, 152

\bibitem[Rich(2013)]{rich13} Rich, R.~M.\ 2013, Planets, Stars and
  Stellar Systems.~Volume 5: Galactic Structure and Stellar
  Populations, Vol. 5, 271
  
\bibitem[Rojas-Arriagada et al.(2014)]{rojas14} Rojas-Arriagada, A.,
  Recio-Blanco, A., Hill, V., et al.\ 2014, \aap, 569, A103

\bibitem[Ryde et al.(2016)]{ryde16} Ryde, N., Schultheis, M., Grieco,
  V., et al.\ 2016, \aj, 151, 1

\bibitem[Scannapieco \& Tissera(2003)]{scannapieco03} Scannapieco, C.,
  \& Tissera, P.~B.\ 2003, \mnras, 338, 880

\bibitem[Stetson(1987)]{stetson87} Stetson, P.~B.\ 1987, \pasp, 99, 191 

\bibitem[Stetson(1994)]{stetson94} Stetson, P.~B.\ 1994, \pasp, 106, 250 

\bibitem[Tsujimoto \& Bekki(2012)]{tsu12} Tsujimoto, T., \& Bekki,
  K.\ 2012, \apjl, 751, L35

\bibitem[Ubeda \& Anderson (2012)]{ubedajay} Ubeda, L., Anderson, J.,
  STScI Inst. Sci. Rep. ACS 2012-03 (Baltimore: STScI)

\bibitem[Valenti et al.(2007)]{valenti07} Valenti, E., Ferraro, F.~R.,
  \& Origlia, L.\ 2007, \aj, 133, 1287

\bibitem[Valenti et al.(2013)]{valenti13} Valenti, E., Zoccali, M.,
  Renzini, A., et al.\ 2013, \aap, 559, A98

\bibitem[Valenti et al.(2016)]{valenti16} Valenti, E., Zoccali, M.,
  Gonzalez, O.~A., et al.\ 2016, \aap, 587, L6

\bibitem[VandenBerg et al.(2015)]{van15} VandenBerg, D.~A., Stetson,
  P.~B., \& Brown, T.~M.\ 2015, \apj, 805, 103

\bibitem[Zanella et al.(2015)]{zanella15} Zanella, A., Daddi, E., Le
  Floc'h, E., et al.\ 2015, \nat, 521, 54
 
\end{thebibliography}
\end{document}